\documentclass{article}

\usepackage{arxiv}

\usepackage[utf8]{inputenc} % allow utf-8 input
\usepackage[T1]{fontenc}    % use 8-bit T1 fonts
\usepackage{hyperref}       % hyperlinks
\usepackage{url}            % simple URL typesetting
\usepackage{booktabs}       % professional-quality tables
\usepackage{amsfonts}       % blackboard math symbols
\usepackage{nicefrac}       % compact symbols for 1/2, etc.
\usepackage{microtype}      % microtypography
\usepackage{lipsum}
\usepackage{graphicx}
\usepackage{float}
\usepackage{amsmath}
\usepackage{hyperref}
\usepackage{verbatim}
\usepackage{multirow}
\graphicspath{ {./images/} }

\title{Med-TTT: Vision Test-Time Training model for Medical Image Segmentation}

\author{
 Jiashu Xu \\
  School of Science\\
  Harbin Institute of Technology, Shenzhen\\
  \texttt{jiashu.xu04@gmail.com} \\
}

\begin{document}
\maketitle
\begin{abstract}
Medical image segmentation plays a crucial role in clinical diagnosis and treatment planning. Although models based on convolutional neural networks (CNNs) and Transformers have achieved remarkable success in medical image segmentation tasks, they still face challenges such as high computational complexity and the loss of local features when capturing long-range dependencies. To address these limitations, we propose Med-TTT, a visual backbone network integrated with Test-Time Training (TTT) layers, which incorporates dynamic adjustment capabilities. Med-TTT introduces the Vision-TTT layer, which enables effective modeling of long-range dependencies with linear computational complexity and adaptive parameter adjustment during inference. Furthermore, we designed a multi-resolution fusion mechanism to combine image features at different scales, facilitating the identification of subtle lesion characteristics in complex backgrounds. At the same time, we adopt a frequency domain feature enhancement strategy based on high pass filtering, which can better capture texture and fine-grained details in images. Experimental results demonstrate that Med-TTT significantly outperforms existing methods on multiple medical image datasets, exhibiting strong segmentation capabilities, particularly in complex image backgrounds. The model achieves leading performance in terms of accuracy, sensitivity, and Dice coefficient, providing an efficient and robust solution for the field of medical image segmentation.The code is available at \href{https://github.com/Jiashu-Xu/Med-TTT}{https://github.com/Jiashu-Xu/Med-TTT}.
\end{abstract}

% keywords can be removed

\section{Introduction}
Modern medical research is heavily reliant on various types of medical imaging.\cite{litjens2017survey} Medical images aim to provide accurate visual representations of the structure and function of different tissues and organs in the human body, assisting medical professionals and researchers in comprehensively understanding normal and pathological conditions. Whether in cutting-edge laboratory research or in clinical disease diagnosis, the rich information provided by medical image analysis is crucial for making scientific inferences and accurate diagnoses.\cite{chen2022recent} Furthermore, automatic medical image segmentation technology can assist doctors in faster pathological diagnosis, thereby improving patient care efficiency.

Due to their strong feature representation capabilities, Convolutional Neural Networks (CNNs) have been widely applied in the field of medical image segmentation and have achieved promising results\cite{shin2016deep}. Fully Convolutional Networks (FCNs)\cite{sun2021segmentation}, as an advanced version of CNNs, enable pixel-level segmentation of images of any size. With further research, U-Net emerged as a novel approach based on FCNs\cite{ronneberger2015u}, using a symmetric encoder-decoder structure with skip connections to effectively enhance contextual information. Raj et al.\cite{raj2022enhanced} demonstrated that U-Net-based methods could achieve effective segmentation on MRI data, while Safi et al.\cite{safi2023accurate} showed the effectiveness of CNNs in handling brain tumors with varying sizes, locations, and blurred boundaries. These studies underscore the strong performance of CNNs in the field of medical image segmentation, particularly for brain tumor segmentation. 

Although CNN based models have excellent representation capabilities, they exhibit inherent limitations in modeling long-range dependencies within images due to the locality of convolutional kernels. The Transformer model can naturally capture global context, so research interest based on the Transformer model is becoming increasingly strong. TransUnet\cite{chen2021transunet}, a pioneering Transformer-based model, first utilizes the Vision Transformer (ViT) for feature extraction during the encoding stage and then employs CNNs for decoding, demonstrating strong capabilities in capturing global information. TransFuse\cite{zhang2021transfuse} combines ViT and CNN in a parallel architecture, capturing both local and global features simultaneously. In addition, Swin-UNet\cite{cao2022swin} integrates the Swin Transformer into a U-shaped architecture, introducing the first purely Transformer-based U-Net model.

Although Transformer-based models have demonstrated excellent performance in global context modeling, the quadratic complexity of the self-attention mechanism leads to high computational overhead, particularly for dense prediction tasks like medical image segmentation. These challenges prompted us to develop a novel medical image segmentation architecture aimed at capturing robust long-range information while maintaining computational complexity within a manageable range.

In recent years, Test-Time Training (TTT) has emerged as a novel sequence modeling layer, demonstrating linear complexity and highly expressive hidden states\cite{sun2024learning}. TTT treats the traditional fixed hidden state as a model that can be dynamically updated through machine learning and optimized using a self-supervised learning mechanism. This dynamic adjustment enables the model to fine-tune its parameters based on test data, thereby exhibiting greater flexibility and expressive power in capturing complex long-range dependencies. Compared to Transformers and Mamba, TTT layers maintain efficiency while demonstrating superior performance in handling long-context sequences.

In medical image segmentation tasks, targets often exhibit considerable diversity in shape and size\cite{hesamian2019deep,ma2024segment}. Consequently, it is of paramount importance to ensure that these variations are effectively captured, as this significantly impacts the performance of the model. Conventional single-resolution extraction techniques often fail to effectively capture this variability. Multi-resolution features can capture information at various levels: high-resolution branches focus on details such as edges and textures, while low-resolution branches emphasize global structure and context. By combining these multi-scale features, the model can gain a more comprehensive understanding of target regions in complex medical images. Additionally, multi-resolution fusion enables the model to effectively integrate information across different branches, thereby improving segmentation precision. This capability is especially valuable when handling small lesions or complex structures, as multi-level feature fusion significantly enhances the model’s ability to identify important regions.

Moreover, the majority of extant models are predominantly spatial in nature, which may result in the inadvertent omission of crucial information present within the frequency domain. Information in the frequency domain can reveal periodic patterns and global characteristics of images, thereby facilitating the identification of nuances that may prove challenging to discern in the spatial domain\cite{huang2023fvfsnet,hai2022combining}. These include textures, edges, and periodic structures. This information is of paramount importance for the accurate capture of subtle differences within images.

In this paper, we introduce Med-TTT, a model that integrates the Vision-TTT backbone network, aiming to overcome the limitations of long-range dependency modeling in biomedical image segmentation tasks. The Vision-TTT layer enables dynamic parameter adjustment during testing, making it more effective in capturing both local details and long-range dependencies. Additionally, we incorporated a multi-resolution fusion mechanism and high-pass filtered frequency domain information. Extensive experiments on multiple medical imaging datasets demonstrate that Med-TTT achieves a good balance between model complexity and performance, validating its effectiveness and practicality for medical image segmentation tasks.

In summary, our contributions are as follows:
\begin{itemize}
    \item We introduced Vision-TTT, a visual backbone network integrated with Test-Time Training (TTT) layers, which enables long-range dependency modeling with linear computational complexity while performing self-supervised adaptation during testing. This hybrid design effectively addresses the challenges of long-term dependency modeling and enhances the model’s generalization ability across diverse data distributions.
    \item We incorporated a multi-resolution fusion mechanism and frequency domain information through high-pass filtering. The inclusion of these two mechanisms brings more feature diversity to the TTT-based Vision-TTT, significantly boosting the model’s performance.
    \item We conducted extensive performance evaluations of the proposed model. The results demonstrate that our model achieves high accuracy (96.07\%), mIoU (78.83\%), and DSC (88.16\%), validating its effectiveness.
\end{itemize}

\section{Preliminaries}
\subsection{Theoretical Equivalence}
Consider the linear complexity TTT layer model, with a learning rate of \( \eta = \frac{1}{2} \) using batch gradient descent, and an initial weight of \( W_0 = 0 \). 
The loss function is defined as:
\begin{equation}
\ell(W; x_t) = \left| W \theta_K x_t - \theta_V x_t \right|^2.
\end{equation}
Taking the derivative with respect to \( W \):
\begin{equation}
\nabla_W \ell(W; x_t) = 2 \left( W \theta_K x_t - \theta_V x_t \right) \left( \theta_K x_t \right)^T.
\end{equation}
Substituting \( W = W_0 = 0 \), we have:
\begin{equation}
\nabla \ell(W_0; x_t) = -2 \left( \theta_V x_t \right) \left( \theta_K x_t \right)^T.
\end{equation}
Using the update formula for \( W_t \):
\begin{equation}
W_t = W_{t-1} - \eta \nabla \ell(W_0; x_t) = W_0 - \eta \sum_{s=1}^{t} \nabla \ell(W_0; x_s) = \sum_{s=1}^{t} \left( \theta_V x_s \right) \left( \theta_K x_s \right)^T.
\end{equation}
Thus, the output \( z_t \) becomes:
\begin{equation}
z_t = f \left( \theta_Q x_t; W_t \right) = \sum_{s=1}^{t} \left( \theta_V x_s \right) \left( \theta_K x_s \right)^T \left( \theta_Q x_t \right),
\end{equation}
which is equivalent to linear self-attention.

For non-parametric models \( f \), since \( w_t \) does not exist, we use the notation \( f(x; x_1, \ldots, x_t) \). Consider the TTT layer using the Nadaraya-Watson estimator, defined as:

\begin{equation}
f(x; x_1, \ldots, x_t) = \frac{\sum_{s=1}^{t} \kappa(x, x_s) y_s}{\sum_{s=1}^{t} \kappa(x, x_s)},
\end{equation}

where \( y_s = \theta_V x_s \), and \( \kappa(x, x'; \theta_K, \theta_Q) \propto e^{(\theta_K x)^T (\theta_Q x')} \) is a kernel function with parameters \( \theta_K \) and \( \theta_Q \).

Substituting the kernel function into \( f \), we obtain:

\begin{equation}
f(x; x_1, \ldots, x_t) = \frac{\sum_{s=1}^{t} \exp \left( (\theta_K x)^T (\theta_Q x_s) \right) y_s}{\sum_{s=1}^{t} \exp \left( (\theta_K x)^T (\theta_Q x_s) \right)}.
\end{equation}

Since \( \text{softmax}(x) = \frac{\exp(x)}{\sum \exp(x)} \), it can be observed that this is equivalent to the self-attention output:

\begin{equation}
z_t = \sum_{s=1}^{t} \text{softmax} \left( \frac{(\theta_K x_s)^T (\theta_Q x_t)}{\sqrt{d_k}} \right) y_s.
\end{equation}

These equivalence demonstrates that the TTT layer can be considered a generalization of the self-attention mechanism, allowing it to not only match existing performance but also introduce new flexibility for handling non-parametric learning tasks, such as non-parametric models and mini-batch gradient descent in various learning methods.

\subsection{Linear Complexity Implementation}
The TTT layer introduces a mini-batch gradient descent method. In standard online gradient descent, each time step’s hidden state is sequentially updated. To achieve parallelism, the TTT layer utilizes mini-batches to compute gradients and update hidden states simultaneously. Specifically, the input is divided into mini-batches of size \( K \times K \), with a total of ${\frac{H}{K}}\times{\frac{W}{K}}$ mini-batches. The gradient calculation within each mini-batch can be parallelized, resulting in a computational complexity of \( O(K^2) \) for a single mini-batch. Consequently, the overall complexity is:

\begin{equation}
O\left(\frac{H}{K}\times\frac{W}{K}\times K^2\right)=O(H\times W)=O(N)
\end{equation}

where \( N \) represents the total number of pixels in the image. In this way, the TTT layer reduces the traditional quadratic complexity (e.g., the self-attention mechanism in Transformers) to linear complexity. The hidden states are updated through parallel computation, significantly enhancing efficiency. Additionally, the TTT layer confines the computation within each mini-batch, further lowering the overall computational cost. This optimization allows the TTT layer to more effectively handle long-range contextual information while maintaining efficient sequence modeling, even under limited computational resources.

\section{Method}
\begin{figure}[htbp]
    \centering
    \includegraphics[width=1\linewidth]{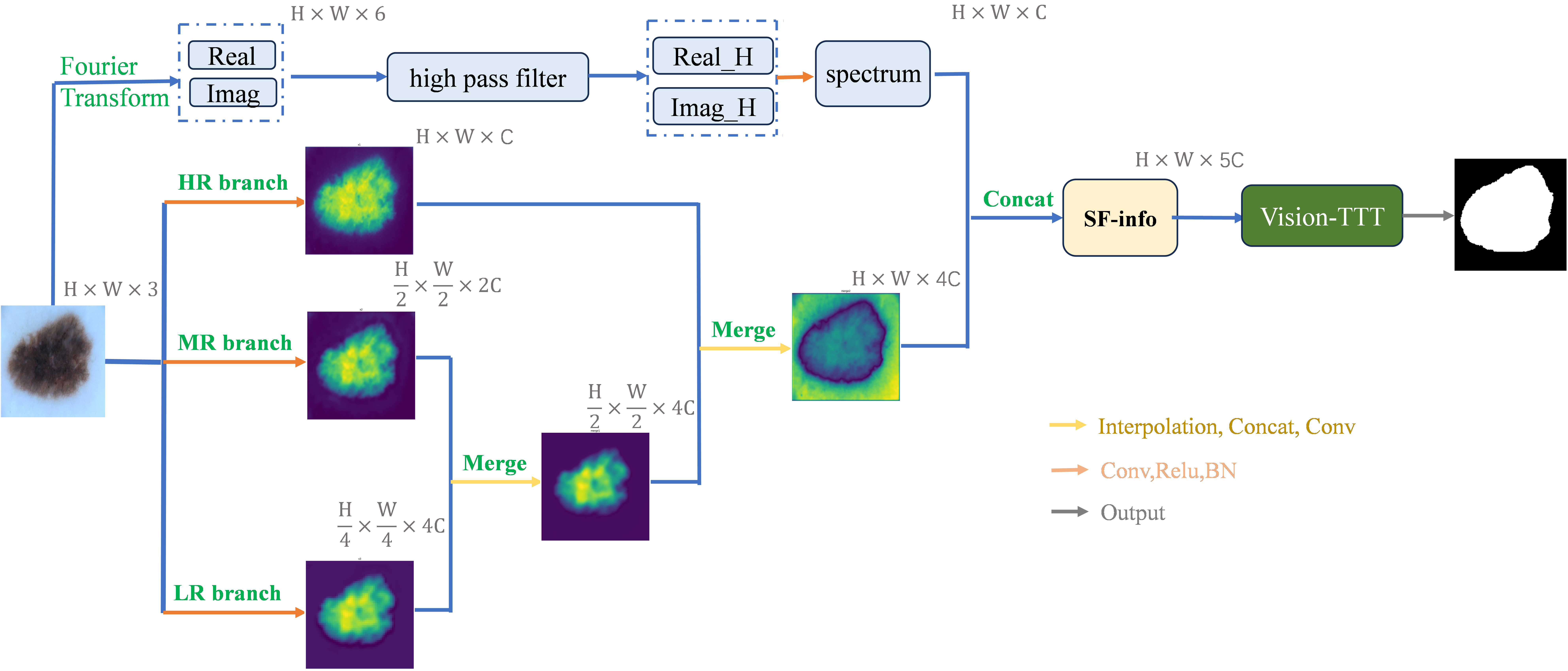}
    \caption{Model structure}
    \label{fig:model-structure}
\end{figure}

The structure of Med-TTT is shown in fig \ref{fig:model-structure}. Med-TTT aims to achieve more accurate medical image segmentation by integrating multi-resolution fusion, frequency domain information, and the Vision-TTT layer based on TTT. 

\subsection{Vision-TTT Layer}
Traditional sequence modeling methods, such as Recurrent Neural Networks (RNNs), compress the context of a sequence input \(x_1, \ldots, x_t\) into a fixed-size hidden state \( h_t \). In RNN-based models, the hidden state \( h_t \) at time step \( t \) is updated based on the hidden state \( h_{t-1} \) from the previous time step and the current input \( x_t \) using linear transformation matrices \( \theta_h \) and \( \theta_x \), along with a non-linear activation function \( \sigma \):$h_t = \sigma(\theta_h h_{t-1} + \theta_x x_t)$, the output \( z_t \) is then generated from the hidden state:$z_t = \phi(h_t)$.

Although such methods have achieved some success in handling sequential data, the fixed-size hidden state limits the model’s ability to capture long-range dependencies, especially when the sequence length exceeds the effective representation capacity of the hidden state. To overcome this limitation, we introduce the Vision-TTT layer, a dynamically updated sequence modeling layer.

In the Vision-TTT layer, the hidden state at time step $t$ is treated as a trainable model $f$ with a weight matrix $W_t$ that is updated based on the current input $X_t$:
\begin{equation}
W_{t}=W_{t-1}-\eta\nabla\mathcal{L}(W_{t-1},x_{t})
\end{equation}
where $\eta$ is the learning rate, and $L$ is a self-supervised loss function used to optimize the model’s performance based on the current input. After updating the weight $W_t$, the output token $z_t$ is generated as: $z_t= f(x_t; W_t )$

This dynamic update process allows the model to better capture contextual information and adjust the hidden states over time, significantly enhancing its ability to model long-range dependencies.

In the simplest form of the TTT layer, the self-supervised learning task is achieved by minimizing the error between the original input $X_t$ and its reconstructed version $f(x_t; W_t )$:
\begin{equation}
\mathcal{L}(W;x_{t})=\parallel\theta_{k}f(\tilde{x}_{t};W)-\theta_{v}x_{t}\parallel^2
\end{equation}
Although this simple reconstruction method is effective in some cases, it fails to capture the rich contextual information in complex dependencies, limiting the model’s understanding of the input.

To address this issue, Vision-TTT introduces a more sophisticated self-supervised loss function by utilizing multiple views of the input data. Instead of directly reconstructing the corrupted input, the Vision-TTT layer projects the input into different views using two transformation matrices \(\theta_k\) and \(\theta_v\). View \( K = \theta_k x_t \) captures the key information needed for learning, while the label view \( V = \theta_v x_t \) provides the optimization target for the model. The corresponding loss function becomes
\begin{equation}
L(W; x_t) = \left\| \theta_k f(x_t; W) - \theta_v x_t \right\|^2.
\end{equation}
This design enables the model to focus on critical features of the input, thereby enhancing its ability to capture complex relationships and long-range dependencies within the data.

Once the weight \( W_t \) has been updated, the output token \( z_t \) is generated using the following formula:$z_t = f(\theta_q x_t; W_t)$,where \(\theta_q\) is a transformation matrix used to extract key information from the input, and \( f \) is a parametric function, specifically a multi-layer perceptron (MLP) in this model. The use of \(\theta_q\) introduces additional flexibility during inference, allowing the model to emphasize the most informative features in the context.
\begin{figure}[htbp]
    \centering
    \includegraphics[width=0.75\linewidth]{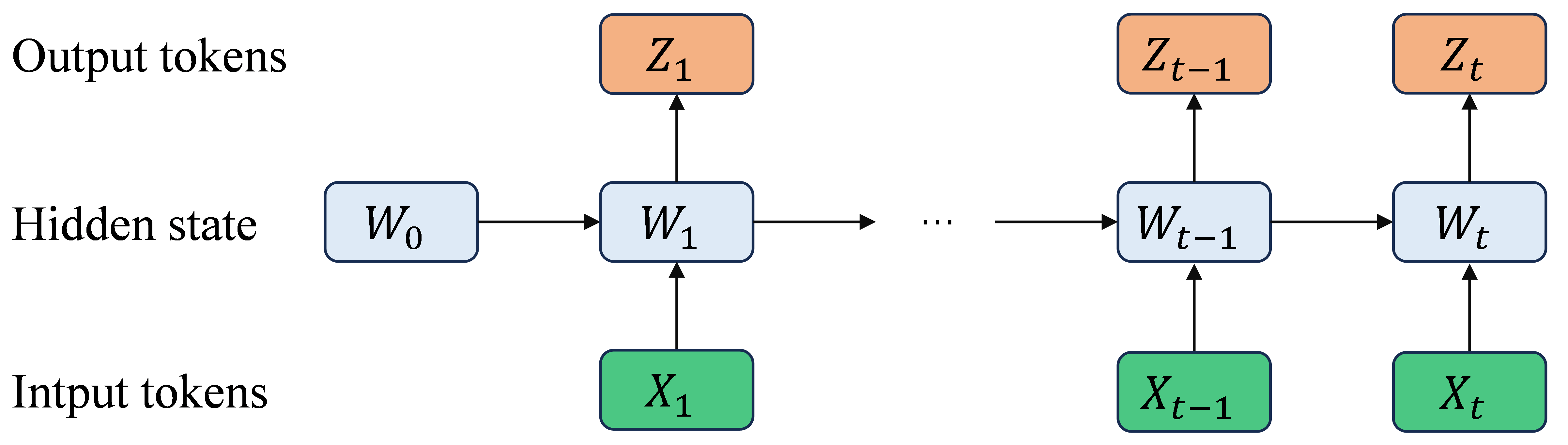}
    \caption{Schematic diagram of the ttt-layer model}
    \label{fig:ttt-layer}
\end{figure}

\subsection{Multi-Resolution Fusion and Frequency Domain Information Integration}
We introduce a multi-resolution fusion mechanism in Med-TTT that processes the input image through three separate branches: high-resolution, medium-resolution, and low-resolution. The high-resolution branch extracts fine-grained features while maintaining the original resolution of the input image; the medium-resolution branch captures deeper features through downsampling, reducing the image size; and the low-resolution branch further downscales the image, capturing broader contextual information. During feature fusion, information is combined from low to high resolutions.

The multi-resolution fusion mechanism enables the model to effectively capture details at different scales, ensuring that no critical information is lost when identifying small lesions or structures in medical images. The low-resolution branch, in particular, provides a more global perspective, allowing the model to better handle complex backgrounds in medical imaging.

Capturing both fine details and distinguishing between background and foreground is critical for medical image segmentation. Therefore, we introduce frequency domain information through high-pass filtering to enhance the model’s ability to identify important features in the image. First, we apply a 2D Fourier Transform to the input image to convert it into the frequency domain, resulting in real and imaginary parts. By calculating the frequency magnitude, we can identify high-frequency components, which often correspond to edges, textures, and detailed information in the image. Next, a high-pass filter is applied to retain only the high-frequency components, filtering out low-frequency background noise, thus extracting features crucial for the segmentation task.

By incorporating high-pass filtered frequency domain information, the details and edge features in the image are enhanced, helping the model to more accurately identify lesion areas during segmentation. Moreover, integrating high-frequency information with features from other branches further enriches the model’s feature representation, enabling it to capture subtle but significant variations in complex medical images.

\subsection{Loss function}
Med-TTT is proposed as a means of obtaining more stable results and higher accuracy in medical image segmentation tasks. Therefore, we employ the Diceloss and Cross-entropy loss function, which are the most elementary in medical image analysis. We then combine these two elements. By calculating the loss at the batch level, we have devised a loss function that can mitigate the fluctuations in loss observed for individual samples, which may be attributable to random noise or misclassification of minor structures. This enables the model to converge in a more stable manner. Furthermore, it can enhance the overall segmentation performance, particularly in scenarios where data distribution is uneven, and facilitate more accurate weighting of different samples. The specific loss function is presented in \eqref{equ:loss}.$\alpha$ refer to the weights of loss functions, which is set to $0.5$ by default.
\begin{equation}
\label{equ:loss}
\left\{
\begin{aligned}
\text{Loss}_{\text{B}} &= (1-\alpha)\times\text{CE}_{\text{B}}+\alpha\times\text{Dice Loss}_{\text{B}} \\
\mathrm{CE}_{\mathrm{B}} &= -\frac{1}{B}\sum_{b=1}^{B}\text{Target}_{b}\log(\text{Input}_{b}) \\
\text{Dice loss}_{\text{B}} &= 1-\frac{2\sum_{b=1}^{B}\mathrm{Input}_{b}^{}\times\mathrm{Target}_{b}^{}}{\sum_{b=1}^{B}\left(\mathrm{Input}_{b}^{}+\mathrm{Target}_{b}^{}\right)+\epsilon}
\end{aligned}
\right.
\end{equation}

\section{Experiment}
\subsection{Datasets}
We conduct comprehensive experiments on Med-TTT for medical image segmentation tasks. Specifically, we evaluate the performance of Med-TTT on medical image segmentation tasks on the ISIC17 and ISIC18 datasets.
\begin{itemize}
    \item \textbf{ISIC2017}:The ISIC2017 dataset contains three categories of diseases, melanoma, seborrheic keratosis, and benign nevus, 2,750 images, ground truth, and category labels. There are 2,000 images in the training set, 150 images in the validation set, and 600 images in the test set, and the color depth of the skin disease images is 24 bits, and the image sizes range from 767×576 to 6,621×4,441. The validation and test sets also include unlabeled hyperpixel images. The category labels are stored in tables and the datasets need to be preprocessed before training the model.
    
    \item \textbf{ISIC2018}:The ISIC2018 dataset contains different numbers of disease images for classification and segmentation, for the segmentation task, a total of 2,594 images were used as the training set, and 100 and 1,000 images were used as the validation and test sets, respectively. For the classification task, a total of 12,500 images were included, of which the training set contained a total of 10,015 images of 7 categories of diseases, namely actinic keratoses (327), basal cell carcinoma (514), benign keratoses (1,099), dermatofibromas (115), melanomas (1,113), melanocytic naevi (6,705), and vascular skin lesions (142). The seven classes of images in the classification task dataset are mixed in the same folder, and the labels are stored in tables that require preprocessing.
\end{itemize}
\subsection{Comparison with SOTA models}
We compare Med-TTT with some state-of-the-art models and some recent mamba-based model, presenting the experimental results in Table \ref{tab:isic}. 

For the ISIC2017 and ISIC2018 datasets, Med-TTT performs well on mIoU and Dice compared to other models. Specifically, Med-TTT has a 1.01\% and 0.78\% advantage over HC-Mamba on mIoU and Dice, respectively, while it has a 1.85\% and 2.17\% advantage over Unet on mIoU and Dice.
\begin{table}[!th]
	\setlength\tabcolsep{3pt}
	\renewcommand\arraystretch{1.25}
	%\scriptsize
	\caption{Comparative experimental results on the ISIC18 dataset. (\textbf{Bold} indicates the best.)}
	\begin{center}
		\begin{tabular}{c|c|ccccc}
			\hline
			\textbf{Dataset} &\textbf{Model}          & \textbf{mIoU(\%)$\uparrow$}  & \textbf{DSC(\%)$\uparrow$}   & \textbf{Acc(\%)$\uparrow$}   & \textbf{Spe(\%)$\uparrow$}   & \textbf{Sen(\%)$\uparrow$}   \\ \hline
			\multirow{5}{*}{ISIC17} &UNet\cite{ronneberger2015u}  &76.98 &85.99 &94.65 &97.43 & 86.82  \\
			&UTNetV2\cite{utnetv2}                & 76.35          & 86.23          & 94.84          & 98.05          & 84.85          \\
			&TransFuse\cite{zhang2021transfuse}               & 77.21          & 86.40          & 95.17          & 97.98          & 86.14 \\
            &MALUNet\cite{malunet}  &76.78 &87.13 &95.18 &\textbf{98.47} &84.78 \\
			&VM-UNet\cite{ruan2024vmunet} &77.59 &87.03 & 95.40 &97.47 &86.13   \\
                &HC-Mamba\cite{xu2024hc} & 77.88 &87.38 &95.17 &97.47 &86.99 \\
                 &\textbf{Med-TTT} & \textbf{78.83}& \textbf{88.16}& \textbf{96.07}& 97.86& \textbf{87.21}\\\hline\hline
			\multirow{8}{*}{ISIC18}&UNet\cite{ronneberger2015u}                     & 77.86          & 87.55          & 94.05          & 96.69          & 85.86          \\
			&UNet++ \cite{unet++}                & 76.31          & 85.83          & 94.02          & 95.75          & 88.65          \\
			&Att-UNet \cite{attentionunet}             & 76.43          & 86.91          & 94.13          & 96.23          & 87.60          \\    
			&SANet \cite{sanet}                 & 77.52          & 86.59          & 93.39          & 95.97          & \textbf{89.46} \\
			&VM-UNet\cite{ruan2024vmunet} &77.95 &87.61  &94.13   &96.94&85.23 \\
            &HC-Mamba\cite{xu2024hc} &78.42 & 87.89 & 94.24 & 96.98 & 88.90 \\ 
            &\textbf{Med-TTT} & \textbf{78.59}& \textbf{88.01}& \textbf{94.30}& \textbf{96.98} & 85.95\\
            \hline
		\end{tabular}
		\label{tab:isic}
	\end{center}
\end{table}

\subsection{Ablation experiments}
We conducted ablation experiments on the proposed Med-TTT by systematically evaluating the impact of each individual component on the overall performance of the model, and the results are shown in Table~\ref{tab:comparison_settings}.In Settings I and III, the results are suboptimal due to the removal of the Multi-resolution block(MR-block), which result in a reduction in the ability to extract features. Furthermore, Setting II is less effective in comparison to Med-TTT due to the failure to incorporate frequency information(FFF), which results in its inability to utilize global information and to recognize the crucial information embedded in the frequency domain.
\begin{table}[htbp]
\centering
\caption{Comparison of different settings on the proposed Med-TTT}
\begin{tabular}{lccccc}
\toprule
\textbf{Settings} & \textbf{MR-block} & \textbf{FFF} & \textbf{TTT} & \textbf{mIoU(\%)} & \textbf{Dice(\%)}\\
\midrule
I & - & - & $\checkmark$ & 68.63 & 80.62 \\
II & $\checkmark$ & - & $\checkmark$ & 77.44 & 85.48 \\
III & - & $\checkmark$ & $\checkmark$ & 75.54 & 84.12 \\
\midrule
Med-TTT & $\checkmark$ & $\checkmark$ & $\checkmark$ & 78.83 & 88.16 \\
\bottomrule
\end{tabular}
\label{tab:comparison_settings}
\end{table}

\section{Conclusion}
We propose a new medical image segmentation model Med TTT, which integrates a Test Time Training (TTT) layer, multi-resolution fusion mechanism, and frequency domain information to address the limitations of existing CNN and Transformer based models. The Vision-TTT layer, which allows for dynamic parameter adaptation during testing, effectively models long-range dependencies while maintaining linear computational complexity. Furthermore, the incorporation of multi-resolution fusion and high-pass filtered frequency domain features significantly enhances the model's ability to capture detailed texture and complex patterns in medical images. Extensive experiments on multiple datasets demonstrate that Med-TTT achieves superior performance in terms of accuracy, mIoU, and DSC, particularly in challenging segmentation scenarios. The results validate the effectiveness and robustness of the proposed model, making it a valuable contribution to the field of medical image analysis. Future work will explore the application of Med-TTT to additional medical imaging modalities and further enhance its generalization capabilities.

\bibliographystyle{unsrt}  
\bibliography{references}  %%% Remove comment to use the external .bib file (using bibtex).


\end{document}